\documentclass{PoS}

\PoS{PoS(HEP2005)051}

\title{Inclusive jet cross-sections and dijet azimuthal decorrelations with D\O}

\ShortTitle{Inclusive jet cross-sections and dijet azimuthal decorrelations with D0}

\author{\speaker{Raimund Str\"ohmer}\thanks{For the D\O~Collaboration.}\\
        LMU Muenchen, Germany\\
        E-mail: \email{Raimund.Stroehmer@Physik.Uni-Muenchen.de}}

\usepackage{epsfig}

\abstract{
We present a preliminary measurement of the inclusive jet cross-sections
based on an integrated luminosity of 378$pb^{-1}$ acquired with the 
D\O~detector between 2002 and 2004 at a center of mass energy
of $\sqrt{s}=1.96$ TeV  and a measurement of azimuthal dijet
decorrelations based on an integrated luminosity of 150 $pb^{-1}$.
The cross section measurements are based on an iterative cone algorithm
with a cone size of $R=0.7$. They are performed in two rapidity
bins between 0.0 and 0.8. The measurements are in good agreement with 
next to leading order calculations.

The azimuthal angle between the two leading jets is sensitive to higher
order QCD effects. The measurement of  dijet azimuthal decorrelations therefore
probes these effects without explicitly reconstructing more than two jets.
 Except for large azimuthal angles where soft effects
are important the measurements are well described by the next to leading order
perturbation theory.  
}
\FullConference{International Europhysics Conference on High Energy Physics\\
		 July 21st - 27th 2005\\
		 Lisboa, Portugal}

\begin{document}

\section{Inclusive jet cross-sections}
Both the increased center of mass energy and the higher integrated
luminosity made it possible to measure the inclusive jet cross section 
at higher transverse jet momenta as compared to Run I. 
At  $p_T=500$ GeV for example
the cross section is according to pQCD calculations expected to increase
by 300~\% due to the change of center of mass energy from $\sqrt{s}=1.8$ TeV 
in  Run I to  $\sqrt{s}=1.96$ TeV in Run II.
At the highest jet $p_T$ 
quark anti-quark fusion diagrams dominate but at $p_T=500$ GeV the
quark-gluon diagrams still contribute with about 30\% to the 
cross section and therfore give sensitivity to the gluon
density in the proton.

The jets are reconstructed with an iterative cone algorithm~\cite{con}
with a cone size of $R=0.7$. The inclusion of the midpoints between jets
as additional seeds avoids infrared instabilities of simpler cone algorithms.
\begin{figure}
\begin{center}
\epsfig{file=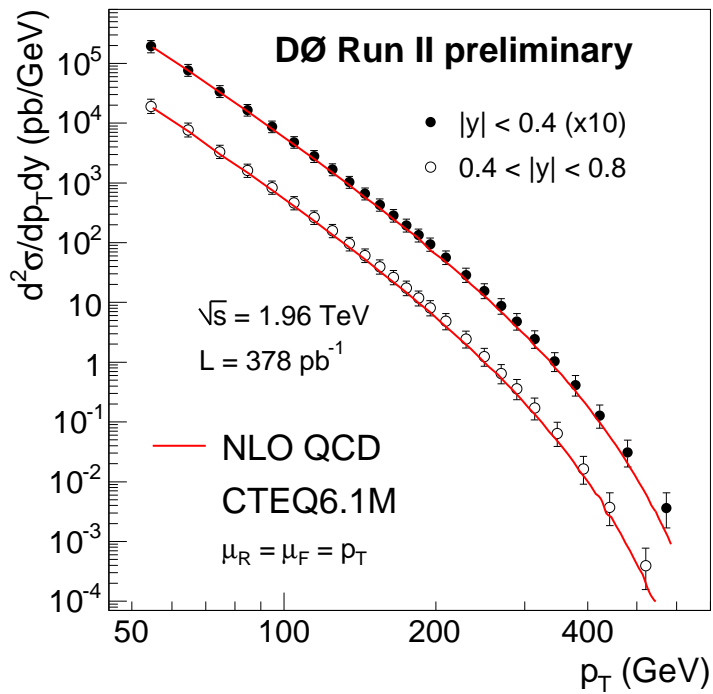,width=10cm} 
\vspace*{.2cm}
\end{center}
\caption{The inclusive jet cross section in two regions of jet rapidity.
The error bars indicate the total experimental error. The predictions
from NLO pQCD are overlaid on the data.
 \label{fig-1}}
\end{figure}

\begin{figure}
\begin{center}
\epsfig{file=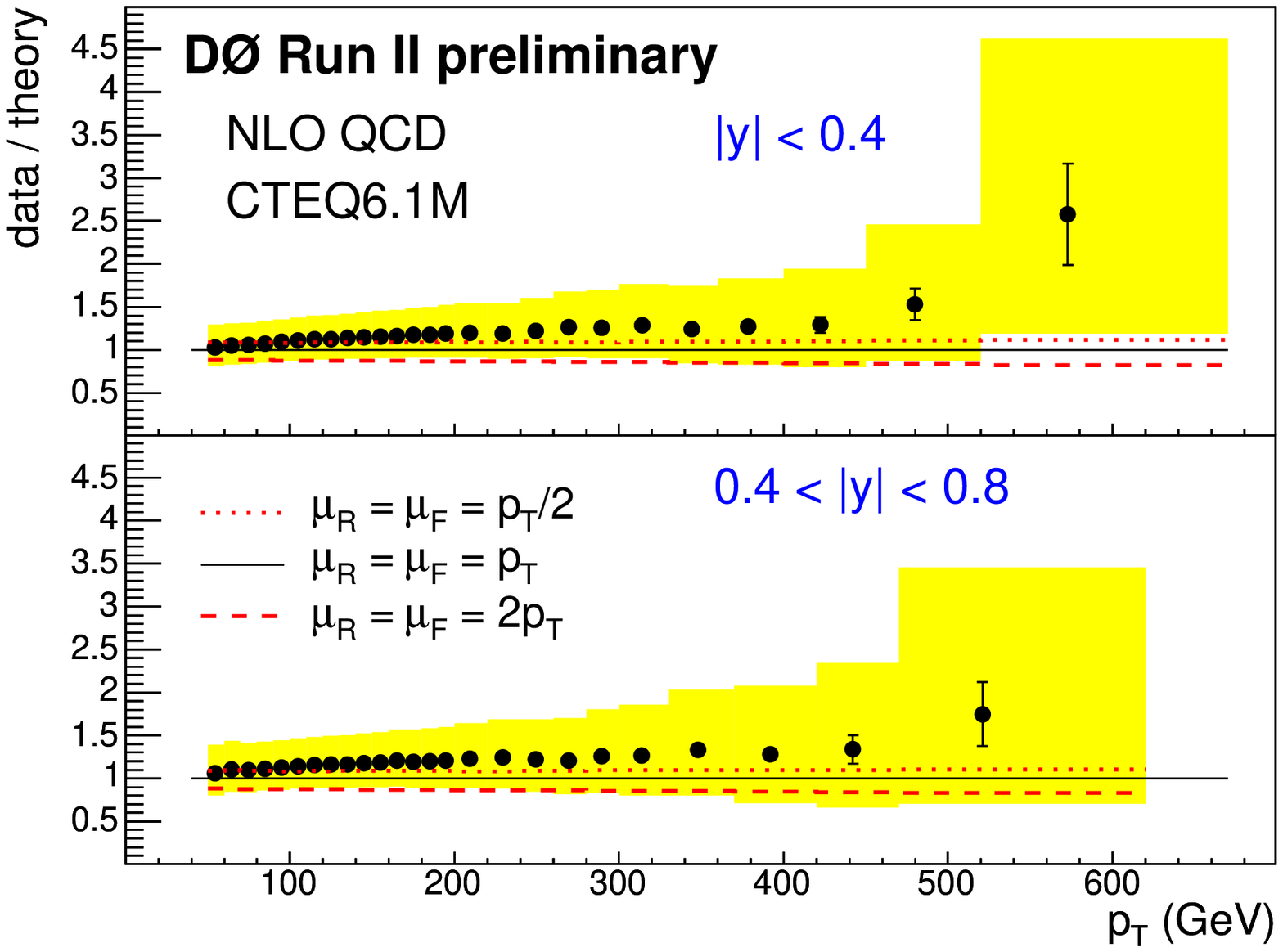,width=7.5cm} 
\epsfig{file=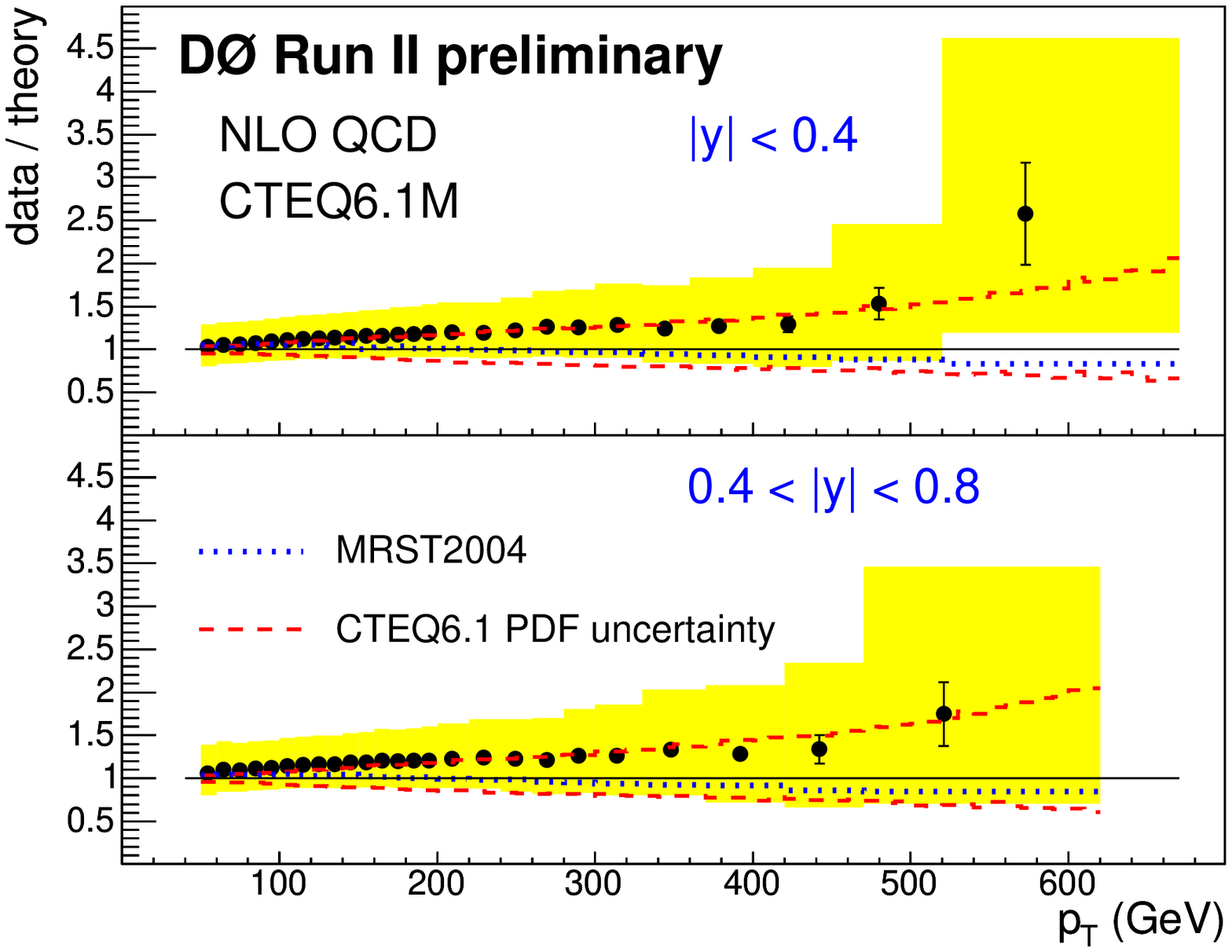,width=7.5cm} 
\end{center}
\caption{Ratio of inclusive jet cross section to the predictions
from NLO pQCD. The bands indicate the total experimental error.
The dashed and dotted lines indicate uncertainties  
due to the renormalisation and factorization scale  
 and uncertainties due to the PDF's.
\label{fig-2}}
\end{figure}

The dominant systematic uncertainty is due to the understanding of the 
jet energy calibration. Due to the steeply falling cross section spectrum 
shifts in the energy scale cause large effects on the measured cross section.
The jet energy scale is estimated from the $p_T$ inbalance in photon
plus jet events~\cite{pho}. 
The effects of the underlying event and hadronization are estimated to be
small. Therefore the measured cross sections are only corrected 
for the jet energy 
measurement selection, for efficiencies and for bin migrations due to the $p_T$
resolution. Figure~\ref{fig-1} shows the measurements in two
rapidity bins ($|y|<0.4$ and $0.4<|y|<0.8$). The measurements are
compared with a NLO QCD calculation using the program NLOJET++~\cite{nlo}
and PDFs from CTEQ6.1M~\cite{CTEQ} and MRST2004~\cite{MRST}.   
In Figure~\ref{fig-2} the ratio of the measured and the predicted 
cross section are shown. The band indicates the total experimental 
uncertainty while the dashed and dotted lines  indicate uncertainties  
due to the renormalisation 
and factorization scale and uncertainties due to PDF uncertainties.

\section{Azimuthal decorrelations}
The azimuthal angle between the two highest $p_T$ jets can be used
to study higher order effects without explicitly reconstructing additional
jets. The distribution of the azimuthal angle close to $180^{o}$ is
strongly effected by soft QCD effects. On the other hand large deviations
from the back to back topology are caused by hard parton emissions. Angles
between the two highest $p_T$ jets smaller than  $120^{o}$ can only
be achieved if more than one hard parton is emitted.

\begin{figure}
\begin{center}
\epsfig{file=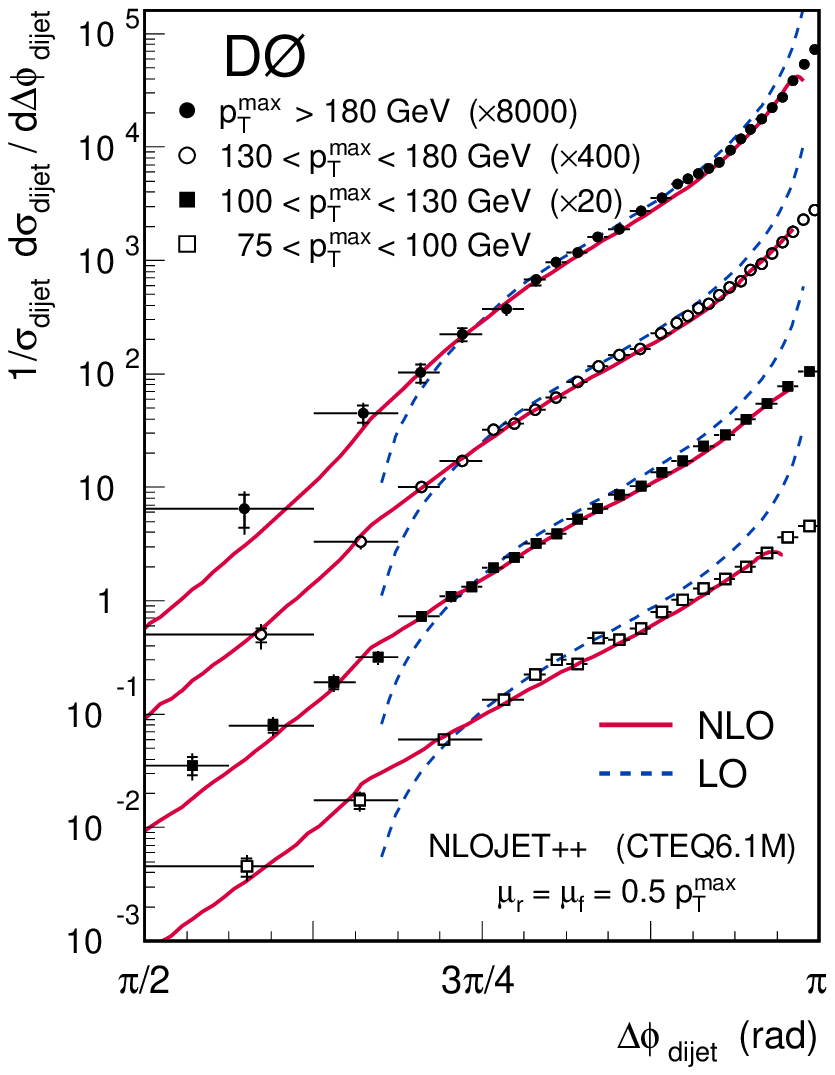,width=7.5cm} 
\epsfig{file=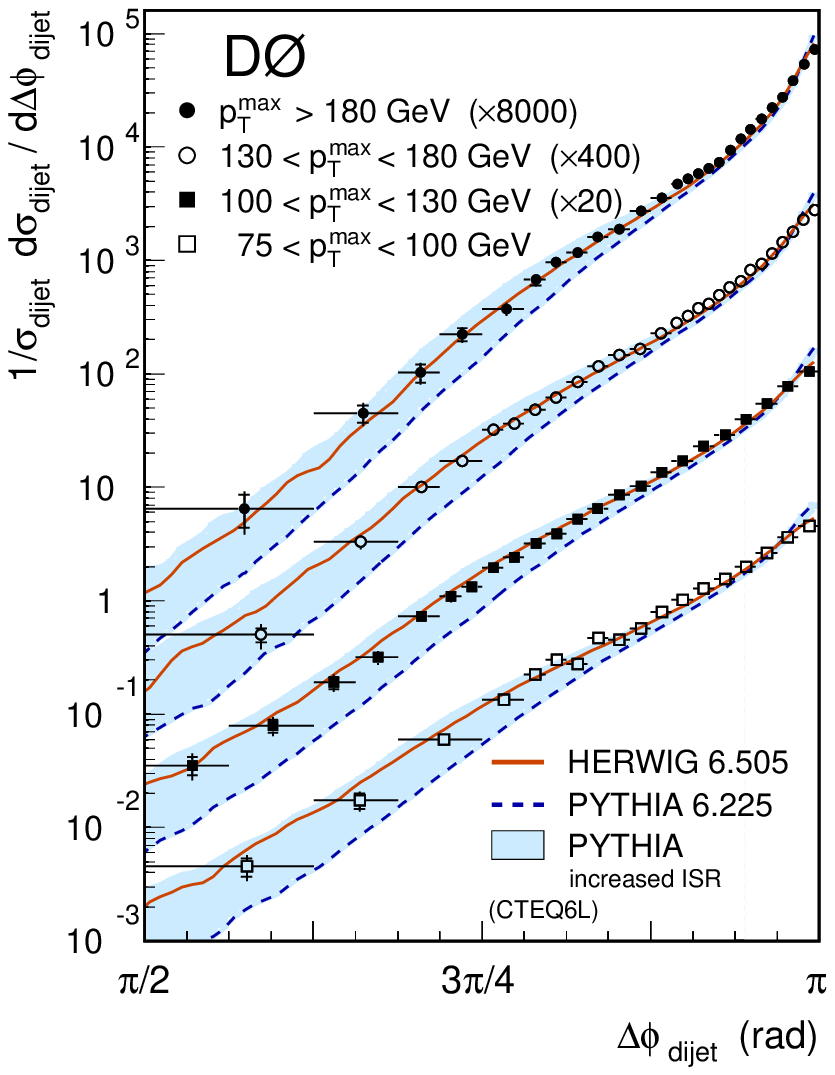,width=7.5cm} 
\vspace*{.2cm}
\end{center}
\caption{
The $\Delta\phi_{dijet}$ distribution in four regions of $p_T^{max}$
compared to NLO an LO pQCD predictions and to the predictions
of QCD shower Monte Carlos.
\label{fig-3}}
\end{figure}

The observable was defined as the differential cross section in $\Delta \phi$
normalized to the cross section integrated over the full $\Delta \phi$ range.
This normalization reduces both theoretical and experimental uncertainty
especially the uncertainty due to the energy scale.
As in the cross section measurement  jets are reconstructed with an 
iterative cone algorithm~\cite{con} with a cone size of $R=0.7$.
The two highest $p_T$ jets are required to have rapidities $|y|<0.5$.
The second highest $p_T$ jet was required to have $p_T>40$ GeV. The
angular distribution is then measured in bins of the $p_T$ of the
highest $p_T$ jet.  
The measurement is compared in Figure~\ref{fig-3} with LO and NLO pQCD
calculations~\cite{nlo,CTEQ} and with the Monte Carlo generators 
HERWIG~\cite{herw} and PYTHIA~\cite{phy}.
For angles close to $180^{o}$ the pQCD calculations become numerically
unstable and are not shown. In this region soft QCD effects are large.
Except for that region the NLO prediction describes the measurements well.
While the leading order prediction can neither describe the region of
small angles (where more than one parton is emitted) nor the region of
large angles.
HERWIG (version 6.505) also describes the data well while PYTHIA with the
default parameters of version 6.225 does not describe the data well.
The agreement can however be greatly improved if the initial state radiation
is increased (the parameter PARP(67) was increased form 1.0 to 4.0).

\end{document}